\documentclass[preprint2]{aastex6}
\usepackage{amsmath,amstext}
\usepackage[english]{babel}
\usepackage{graphicx}
\usepackage{epsfig}
\usepackage{color}
\newcommand{\teff}{$T_\mathrm{eff}$}
\newcommand{\logg}{$\log g$}

\newcommand{\kms}{km\,s$^{-1}$}
\newcommand{\mic}{$\mu \mathrm m$}

\shorttitle{Metal-poor GC giant}
\shortauthors{Ryde et al.}

\begin{document}

\title{Detailed abundance analysis of a metal-poor giant in the Galactic Center}
\author{N. Ryde\altaffilmark{1}, T. K. Fritz\altaffilmark{2}, R. M. Rich\altaffilmark{3}, B. Thorsbro\altaffilmark{1}, M. Schultheis\altaffilmark{4},  L. Origlia\altaffilmark{5}, S. Chatzopoulos\altaffilmark{6}} 
 \email{ryde@astro.lu.se}
\altaffiltext{1}{Lund Observatory, Department of Astronomy and Theoretical Physics, Lund University, Box 43, SE-221 00 Lund, Sweden}
\altaffiltext{2}{Department of Astronomy, University of Virginia, 3530 McCormick Road, Charlottesville, VA 22904, USA}
\altaffiltext{3}{Department of Physics and Astronomy, UCLA, 430 Portola Plaza, Box 951547, Los Angeles, CA 90095-1547, USA}
\altaffiltext{4}{Observatoire de la C\^ote d'Azur, CNRS UMR 7293, BP4229, Laboratoire Lagrange, F-06304 Nice Cedex 4, France}
\altaffiltext{5}{INAF - Osservatorio Astronomico di Bologna, Via Ranzani 1, I-40127 Bologna, Italy}

\altaffiltext{6}{Research Center for Astronomy, Academy of Athens, Soranou Efessiou 4, GR-115 27 Athens, Greece}

\begin{abstract}

We report the first results from our program to examine the metallicity distribution of the Milky Way nuclear star cluster connected to SgrA*, with the goal of inferring the star formation and enrichment history of this system, as well as its connection and relationship with the central 100 pc of the bulge/bar system.  We present the first high resolution (R$\sim 24,000)$, detailed abundance analysis of a $\mathrm{K}=10.2$ metal-poor, alpha-enhanced red giant projected at $1.5$\,pc from the Galactic Center, using NIRSPEC on Keck II. A careful analysis of the dynamics and color of the star locates it at about $26^{+54}_{-16}$\,pc line-of-sight distance in front of the nuclear cluster. It probably belongs to one of the nuclear components (cluster or disk), not to the bar-bulge or classical disk. A detailed spectroscopic synthesis, using a new linelist in the K band, finds [Fe/H]$\sim -1.0$ and [$\alpha$/Fe]$\sim +0.4$, consistent with stars of similar metallicity in the bulge.   As known giants with comparable [Fe/H] and alpha enhancement are old, we conclude that this star is most likely to be a representative of the $\sim 10$ Gyr old population. 
This is also the most metal poor confirmed red giant yet discovered in vicinity of the nuclear cluster of the Galactic Center.  We consider recent reports in the literature of a surprisingly large number of metal poor giants in the Galactic Center, but the reported gravities of $\log g \sim4$ for these stars calls into question their reported metallicities.  

%Metallicities and metallicity gradients in the inner Galactic Bulge and the  nuclear star cluster are %important constraints to structure and formation models of these structures. Not many such measurements have %been made due to the large extinction toward low Galactic latitudes. The few existing measurements show a %narrow spread around approximately solar metallicities. In a project where our aim is to characterize giants %in the nuclear cluster, we have retrieved high signal-to-noise, K-band spectra, observed with  Nirspec at %Keck II. We have observed at a spectral resolution of $R>20,000$, which is needed for retrieving accurate %abundances, through a detailed abundance analysis. Here, we report on the analysis of a K=10.3 giant at 1.6 %pc projected distance from the Galactic Center. A detailed spectroscopic synthesis modelling reveals a %[Fe/H]$\sim-1.0$, alpha-element rich giant. This is the most metal-poor giant discovered yet in the nuclear %cluster in the Galactic Center.
%[Fe/H]$\sim-1.0$This implies...

\end{abstract}

\maketitle
% * <ryde@astro.lu.se> 2016-05-01T15:31:26.896Z:
%
% ^.

\section{Introduction}

The predominance of late-type M giants in the bulge was known since the work of \citet{nassau}  and in fact had marked the bulge as a metal rich, disk population in the 1957 Vatican meeting on stellar populations.  The evolved stellar content was known early on to be very different from that of the halo and globular clusters, even though Baade's (1951)\nocite{baade:51} discovery of RR Lyrae variables in the bulge offered one population in common.  The lack of metal poor giants with [Fe/H] $\sim -1$ was evident from the earliest abundance distributions \citep{rich:88} and was confirmed in all subsequent studies e.g. \citet{fulbright:06}, \citet{zoccali:08}, \citet{johnson:11,johnson:13,johnson:2014}  and \citet{ness:13}.  While aspects of the microlensed dwarf population remain in debate, especially the age distribution, \citet{bensby:13} agree with the giants in finding very few stars with [Fe/H]$<-1$.  These studies all consider bulge fields with $b<-4^o$, although the lack of a metal poor population inward of $b=-4^o$ continues to be confirmed in the GIBS survey \citep{gibs:II}.   Extremely metal poor stars (e.g. [Fe/H]$<-3$) are known in the bulge, but they are so rare that wide field surveys must be undertaken to discover them \citep[see e.g.][]{howes:15} and while they are in the bulge, they are considered to have an origin apart from most of the bulge \citep{koch:16}.   Simply based on an assessment of the nature of the bulge, known abundance gradients in external galaxies, and the well established presence of massive clusters and young stars toward the Galactic Center, it would be expected that few if any metal poor stars might be found within 300 pc of the nucleus.

Indeed, at the Galactic Centre, \citet{ramirez:00}, \citet{carr2000}, \citet{davies2009} by analysing red supergiants  find a metal-rich population narrowly distributed around the Solar metallicity. Similar results are found by  \citet{ryde_schultheis:15} for M giants. Recently though,  \citet{schultheis:15} find the presence of
a metal-poor population beyond 70\,pc from the nuclear cluster, which has a radius of approximately $7\,$pc \citep{fritz16}. \citet{do:15} also report a significant population of metal-poor stars in the nuclear cluster. 

Here we report on the first high-resolution spectroscopy of a  red giant with [Fe/H]$\sim -1$ to be found in the vicinity of the nuclear cluster, at a projected distance of $1.5$\,pc from the Galactic Center. 
We have performed a detailed abundance analysis based on high-resolution, K-band spectra. 
Observing at K band make such investigations possible due to the  much lower extinction at higher wavelengths \citep{cardelli}. The K band extinction toward
the central parsec is only $A_{K_S}=2.74$ with a variation of $\pm0.30$ due
to spatial variations \citep{schoedel10}.
% and toward the Galactic Center regions general values of $A_{K_S}=2.6-2.95$ are found, depending on spatial variability and way of measurement \citep{fritz:11}.

\section{Target selection}

We selected a list of Galactic Center giants from spectra observed with the integral field spectrometer SINFONI \citep{sinfoni,sinfoni2} on the VLT, providing a K-band resolution of $R=4000$ or $R=1500$. The selection for the target group, to which the observed star belongs, was done according to following criteria: a $K_S$ magnitude range of $10<K_S<11$, an angular distance from the Galactic Center of  $R_c(\mathrm{Sgr A^*})>25$\arcsec, and excluding stars with neighbors too close for seeing-limited high resolution spectroscopy. %These criteria 
Initially, SINFONI spectra were only used to ensure qualitatively that the objects are cool in the sense that the CO band-heads exist, but we have not imposed any initial cuts based on the CO band strength or derived effective temperature.
%They were not used for a quantitative temperature based sample selection. 
We also used the catalogs of \citet{blum03,matsunaga09} to exclude some known AGB/LPV stars like Miras, and known red supergiants. The aim of this sample is to develop a relatively unbiased, large sample of old Galactic Center stars that we will use in our study of the metallicity distribution.  A more detailed discussion of our full sample is in preparation.  

\begin{figure}[!tbp]
  \centering
\epsscale{1.00}
\includegraphics[trim={1.5cm 5cm 1cm 2cm},clip,width=1.00\hsize]{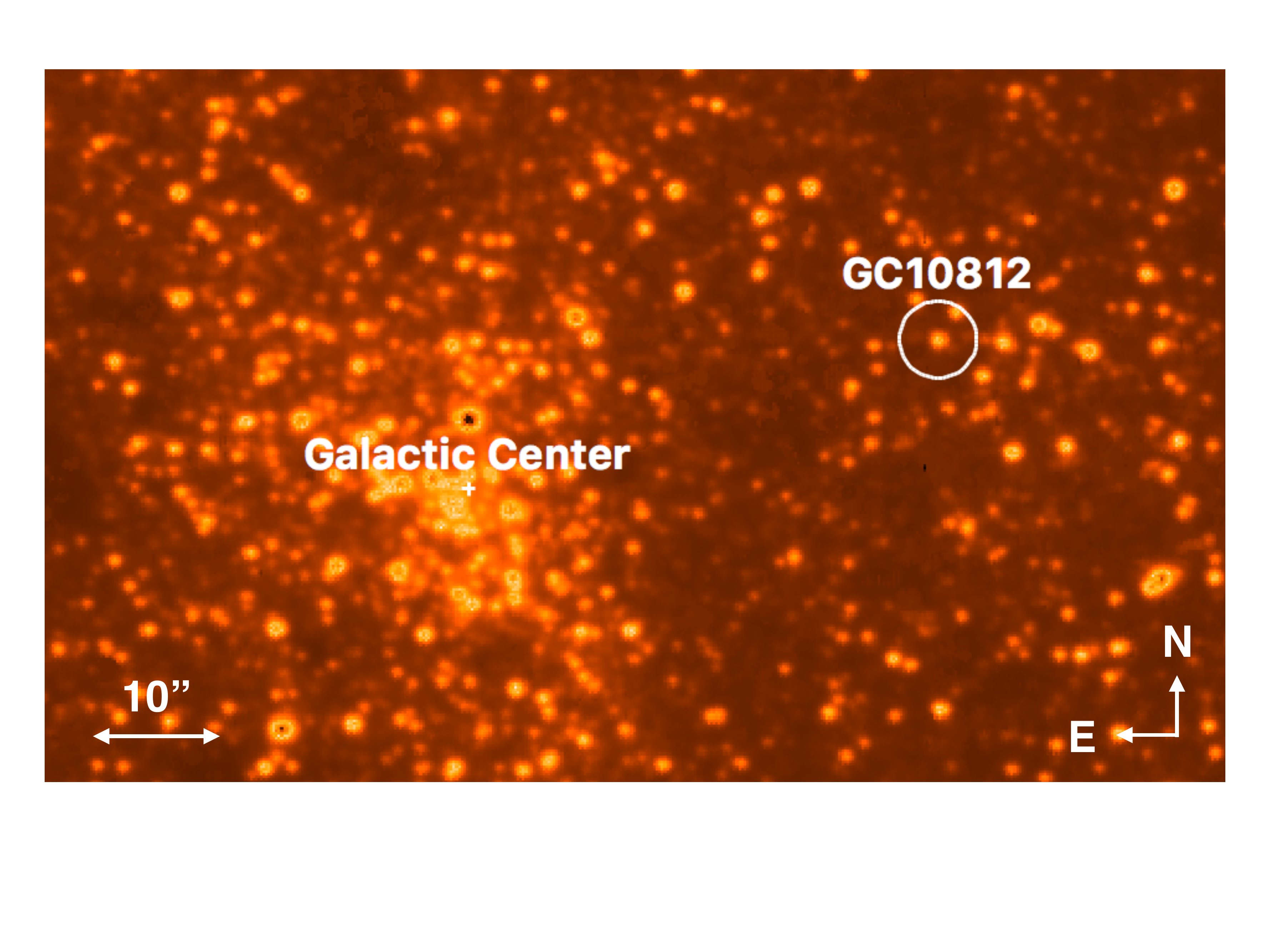} %\includegraphics[angle=-90,width=1.2\hsize]{finding_chart4.pdf}
\caption{UKIDSS finding chart of the Galactic center \citep{ukidss}. GC10812 is marked with a circle. \label{fig:finding}}
\end{figure}

In Figure \ref{fig:finding} we present a finding chart \citep{ukidss} with our observed target, the giant GC10812, indicated (see also Table\,\ref{tab:star}). The star lies at an angular distance of $38.4$\arcsec\ from the Galactic Center.  This corresponds to a projected galactocentric distance of $R_c=1.5$\,pc, adopting the distance to the Galactic Center of $8.3$\,kpc \citep{GCdistance,bland16}.

The $K_S$ magnitude  of GC10812 is $10.25 \pm 0.05$ from \citet{nishiyama2009} and its color is $H-K_S=1.63\pm 0.07$. Using the extinction law of \citet{fritz:11} this leads to an extinction of $A_\mathrm{Ks}=1.94\pm0.10$. The absolute magnitude of GC10812 is $M_{Ks}=-6.48\pm0.12$. This is close to the approximate tip of the RGB with $M_{bol}=-3.5$
(see also \citet{tiede1995}, \citet{omont1999}). Compared to all stars brighter than $M_{Ks}=-5.5$ from \citet{nishiyama2009} within $100"$ of Sgr~A* the star is clearly bluer, $-0.84$ bluer than the median color. That means that only $21$ of $1067$ stars are bluer than GC10812.
There are also extinction variations over the Galactic Center \citep{schoedel10}. To account for them we measure relative color locally using Gemini AO data\footnote{We use H-continuum and K (long) continuum images from program Gs-2013A-Q-15 avoiding data in which the target is saturated.}.
We use stars with $K_S<16$ and account for intrinsic color differences. Dependent on the comparison scale (3 or $6"$) and including other error sources like  scatter  and photometry uncertainty for the target we obtain that GC10812 is $0.75\pm0.10$ bluer than the local median.
Over the full Galactic Center $3.6^{+2.7}_{-1.9}$\% of all stars are that blue or bluer. 
%{\color{blue}(I assume a color difference of $0.1$ mag, now checked, right)} 
The $H-K_S$ color excess corresponds to an extinction difference of $A_\mathrm{Ks}=1.01\pm0.14$ between the $A_\mathrm{Ks}=1.94$ measured for this star and $A_\mathrm{Ks}=2.95$ measured as the average for all stars in this region. 
That is more dust than the $A_\mathrm{Ks}\approx0.4$ which \citet{chatzopoulos15} find within $r=100"$ of Sgr~A*.
However, there is also extinction variation in the plane of sky \citep{schoedel10}, at some places up to $A_\mathrm{Ks}\approx0.8$ were found.
Further, the model of \citet{chatzopoulos15} is very insensitive to dust at greater distances.

To constrain the line of sight position, we use the Galactic Center model of \citet{GCdistance} and derive how much a star needs to be placed in front of Sgr~A* in order for it to be bluer than $3.6^{+2.7}_{-1.9}$\% of its stars. 
We obtain a distance of $D_c=26^{+54}_{-16}$ parsec. %Changed also because before I used r=0 and not the real projected distance
That is in principle a slight underestimate of the distance because the model includes no bar-bulge nor Galactic disc. This contribution is, however, probably small; 
\citet{launhardt02} show that in this part of the nuclear disk the space density due to the bar/bulge is more than 40 times smaller than the contribution due to the nuclear disk. Its absolute extinction also argues for a location within the nuclear disk; an extinction of $A_{Ks}\approx 2$ is unusual for old stars outside the nuclear disk \citep{schultheis14}. 
Whether it is possible that the star still is around the outer rim of the nuclear cluster or not, depends on the precise distance and nuclear cluster definition \citep{schodel:14,GCdistance,fritz16}.
It is however clear that the star is currently within a nuclear component, because the nuclear disk extends to an outer radius of 230 pc \citep{launhardt02}.

\begin{figure}[!tbp]
  \centering
\epsscale{1.00}
\includegraphics[trim={0cm 0cm 0cm 0cm},clip,angle=-90,width=1.00\hsize]{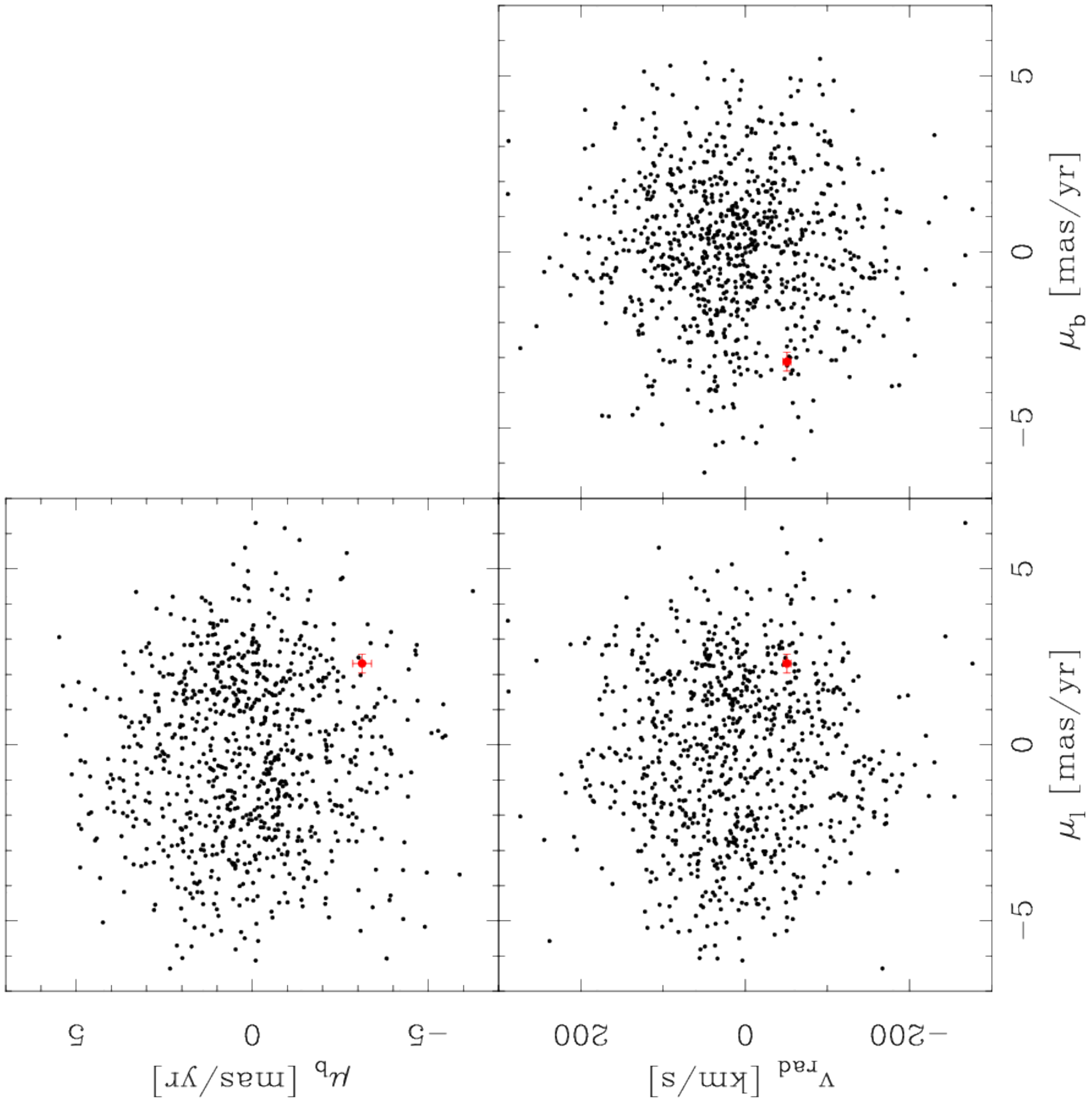} %\includegraphics[angle=-90,w 3.6 pc 3.6 pcidth=1.2\hsize]{finding_chart4.pdf}
\caption{Velocity of GC10812 (red) in comparison with other Galactic Center stars. The other stars are in 0.8 to 3.6 pc projected distance from Sgr~A* \label{fig:velplan}}
\end{figure}

The dynamics measured in \citet{fritz16} are  $\mu_l=2.31\pm0.27$\,mas\,yr$^{-1}$,  $\mu_b=-3.12\pm0.27$\,mas\, yr$^{-1}$, and $v_{rad}=-51\pm5$\,km\,s$^{-1}$ (which we confirm in our measurement of the star's heliocentric velocity from the NIRSPEC high-resolution spectra of $-56.4\,\pm1.0$\,\kms, see Table\,\ref{tab:star}). These velocities are rather typical for a Galactic Centre star, see Figure\,\ref{fig:velplan}. The positive velocity in $l$ and the lower extinction fits as the extinction to a star in front of the Galactic Center \citep{chatzopoulos15}. 

\begin{figure}[!tbp]
  \centering
\epsscale{1.00}
\includegraphics[trim={0cm 0cm 0cm 0cm},angle=-90,width=1.10\hsize]{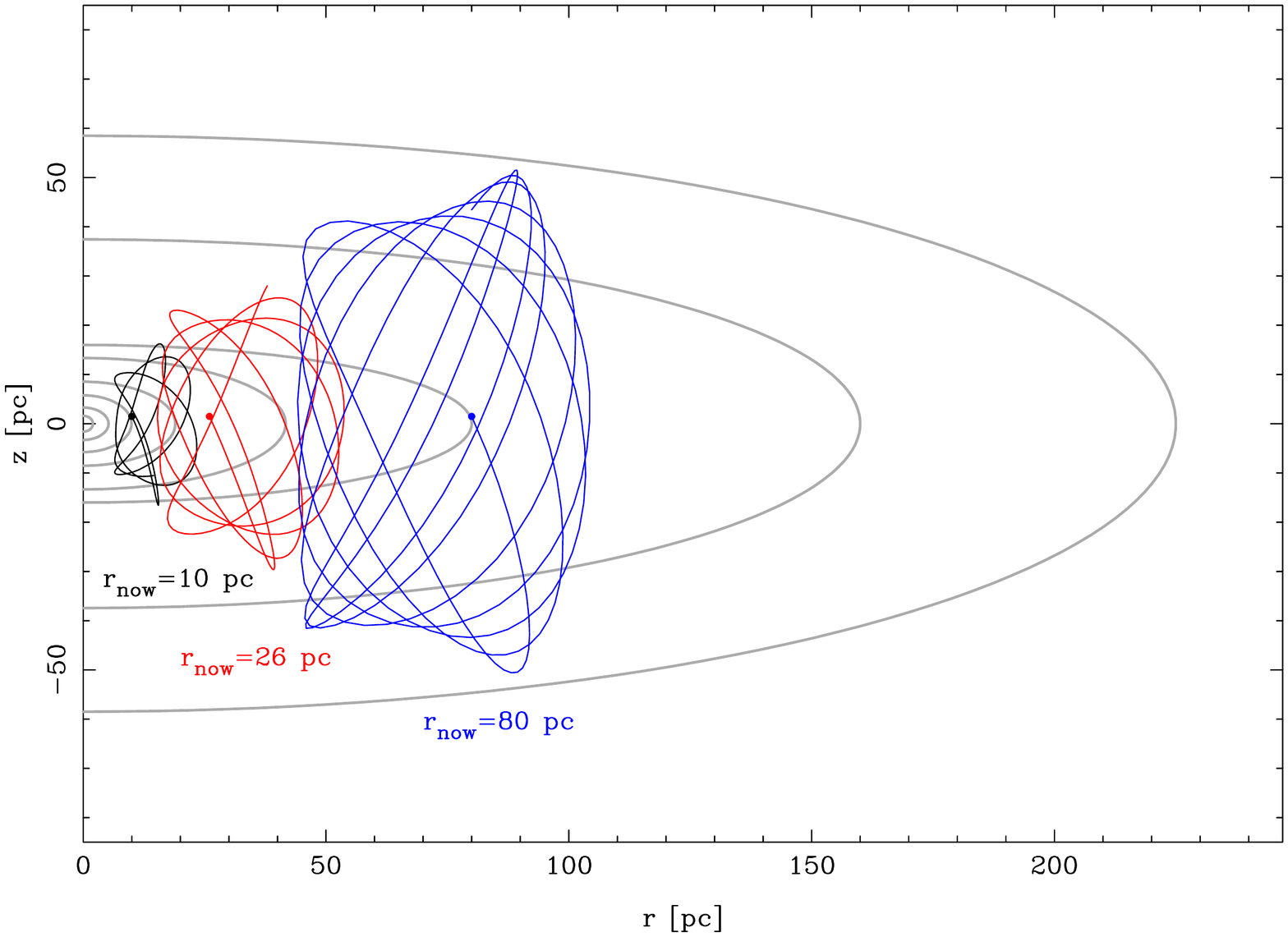} %\includegraphics[angle=-90,width=1.2\hsize]{finding_chart4.pdf}
\caption{Orbit of GC10812. The red curve shows it for the most likely distance in front of the Galactic Center (26 pc), while the other two curves show the orbit when the star would be at the outer/inner edge of the one sigma interval. The gray curves show equal star surface density of contours of the nuclear disk from \citet{fritz16,launhardt02}. The outer most curve marks roughly the outer edge of the nuclear disk. \label{fig:orbits}}
\end{figure}

We calculate the orbit for the stars using the potential of \citet{GCdistance}. This potential includes the supermassive black hole, nuclear cluster, and cluster disk. Three orbits are shown in Figure~\ref{fig:orbits}. For small current distances (black and red orbits in the Figure) the average distance of GC10812 from the Galactic Center is somewhat larger than the current distance from the Galactic Center. That makes membership in the nuclear cluster unlikely, and excludes it nearly certainly if the cluster is assumed to have a Sersic like cutoff, which results in a half light radius of about 5 pc \citep{fritz16,schodel:14}. If it is instead assumed that the outer slope follows a power law  \citep{GCdistance}, a membership in the nuclear cluster is still possible. When the star is currently at large distance (blue orbit in Figure \ref{fig:orbits}), the star does not move much further away on its orbit. We conclude that the star most likely resides within the the nuclear disk and is unlikely to experience any excursions into the bulge or halo.  We conclude based on the distance and kinematics that GC10812 is probably a nuclear disk star.

% \textbf{LIVIA: What is the actual reddening in K estimated for GC10812? 
% I think we should quote our estimated dereddened Ks mag too, and comment whether this star is likely an old giant near the Tip or 
%  a more massive (i.e. younger) giant.
%  Tobias: I added the absolute magnitude before in the section. Somebody else with more knowledge about magnitudes of the populations should use it (together with $T_{eff}$?) to answer the point.}

\section{Observations}

We observed the giant GC10812 on April, 27 2015 with the NIRSPEC spectrometer \citet{nirspec_mclean} mounted on Keck II, at a resolution of $R\sim 24,000$ in the K band, using the $0.432"\times 12"$ slit and the NIRSPEC-7 filter. The retrieved spectra range from 
21,100 to 23,300 \AA, using 5 orders and therefore obtaining about 50\% spectral coverage.  We observed in an ABBA scheme with a nodding throw of 6" on the slit, to achieve proper background and dark subtraction.  A total exposure time on target was $960\,$s. The data were reduced with the NIRSPEC software {\tt redspec} \citep{nirspec} providing final 1-D wavelength-calibrated spectra.  IRAF \citep{IRAF} was subsequently used to normalize the continuum, eliminate obvious cosmic ray hits, and correct for telluric lines (with telluric standard stars). We estimate S/N=90 per pixel in our reduced spectra. A portion of the observed spectrum is shown in Figure \ref{fig:abund}.

%\plotfiddle{PSFILE}{VSIZE}{ROTANG}{HSCALE}{VSCALE}{HTRANS}{VTRANS}
%\plotfiddle{sample.eps}{2.6in}{-90.}{32.}{32.}{-250}{225}

\section{Analysis}

We analyze our spectra, deriving detailed chemical abundances, by calculating synthetic spectra, given the star's fundamental parameters, i.e.  effective temperature (\teff), surface gravity (\logg),  metallicity ([Fe/H]), and microturbulence ($\xi_\mathrm{mic}$) and  a suitable  line list in the K band (see Sect.~\ref{ll})
% Nils: how should we refer to it?

\subsection{The stellar parameters}
The derived $T_\mathrm{eff}=3817\pm 150\,$K, is determined from integrating the strength of the 2.3\mic\ (2-0)CO band in the  SINFONI spectra \footnote{The SINFONI spectrum is from program 087.B-0117.} and using the relation between the CO-band strength and the \teff\ given in \citet{schultheis:16}.  They have shown that this relation works very well in the temperature range between $3200$\,K and $4500$\,K and in the metallicity range between $0.5$\,dex and $-1.2$\,dex with a  typical dispersion of about $150$\,K. %\citet{frogel2001} measured CO bands as well as metallicities from low-resolution spectra %for globular clusters and found
%that the  CO absorption band decreases with decreasing $\rm [Fe/H]$ (see their Fig.~4). %However, this dependency of the CO band on metallicity stars appear
%at about $\rm [Fe/H] < -1.1$. As shown by \citet{schultheis:16} the dependency of
%the \teff\ vs. CO relation is small for $\rm -1.2 < [Fe/H] < 0.5$.

The surface gravity, \logg, is determined photometrically  and assuming a mean distance of $8.3$\,kpc %changed, has that an impact?
to the Galactic Center,  in the same manner as in \citet{ryde_schultheis:15}. A stellar mass of an old star of typically $1\,M_\odot$ is assumed, but the surface gravity is not very sensitive to the mass. A difference in mass of $0.2\,M_\odot$ gives $0.1$\,dex \logg\ difference. In the calculation of the surface gravity we used the extinction law from \citet{fritz:11}, and the bolometric corrections from   \citet{houdashelt2000}. An extinction uncertainty of $A_{Ks}=0.1$ gives $0.2$\,dex error in \logg\ and a bolometric uncertainty of $M_\mathrm{bol}=0.12$ gives $0.15$\,dex uncertainty in \logg. The combined uncertainty is thus $\sqrt{0.1^2 +0.2^2 +0.15^2} = 0.27$\,dex. Our derived surface gravity is thus $\log g = 0.5\pm0.3$ (dex).  

\citet{ramirez2000} found that the combination
of the CO first overtone band together with the Na\,{\sc i} and the Ca\,{\sc i}  lines are sensitive to the surface gravity of the star.
We have therefore also measured
the Na\,{\sc i} and the Ca\,{\sc i} lines in the SINFONI spectra as in \citet{schultheis:16}.  Our derived value for $\log \rm{[EW(CO)/(EW(Na)+EW(Ca))]} = 0.405$ for GC10812 locates our star on the RGB sequence of \citet{ramirez1997}.

We have  chosen a typical value of $\chi_\mathrm{mic}=2.0\pm 0.5$ km\,s$^{-1}$ found in detailed investigations of red giant spectra in the near-IR by \citet{tsuji:08}, see also the discussion in \citet{cunha2007}. 

The location of GC10812 in the Hertzsprung-Russel diagram, using our derived effective temperature, surface gravity, and metallicity (see Table \ref{tab:abund}), is plotted with a big star symbol in Figure\,\ref{fig:cmd}. It fits nicely on the red-giant branch. 

\subsection{Line list} \label{ll}

An atomic line list based on the VALD3 database \citep{vald,vald2,vald3,vald4,vald5} has been constructed (Thorsbro et al. 2017, in prep.). Wavelengths and line-strengths (astrophysical $\log gf$-values) are updated for 575 lines in the K band using the solar center intensity atlas \citep{solar_IR_atlas}. New laboratory measurements of wavelengths and oscillator strengths of Sc lines \citep{pehlivan} and newly calculated oscillator strengths of Mg lines (Pehlivan et al. 2016, in prep) are included.  ABO\footnote{ABO stands for Anstee, Barklem, and O'Mara, authors of the papers describing the theory in \citet{abo1,abo2}} line-broadening theory is included \citep{abo1,abo2} when available. 
% What is ABO standing for ? Nils
In the abundance analysis, we also include molecular line-lists of CN   %(J\o rgensen \& Larsson, 1990)\nocite{jorg_CN}
\citep{jorg_CN} and  CO  \citep{goor}. 

The line list is tested by determining abundances from high-quality spectra of  $\alpha$ Boo \citep{arcturusatlas_II} and 5 thick-disk stars.
The thick-disk giants were observed with NIRSPEC in the same manner as for GC10812. The parameters and abundances of $\alpha$ Boo are from \citet{aboo:param} and of the thick disk stars from the APOGEE pipeline \citep{apogee:14}. The parameter range of these test stars are $4150<T_\mathrm{eff}/\mathrm{K}<4750$, $1.5<\log g< 2.5$, and $-0.5<\mathrm{[Fe/H]}<-0.1$.  We find an excellent agreement, to within $0.05$\,dex, between the abundances we determined from the K band and these reference values. The line list has not yet been tested against super-solar metalicity stars. Thus, the line list can with confidence be used for metal-poor to solar-metallicity cool stars. 

\subsection{Spectral synthesis \label{specsynt}}

We derive our target star's abundances by comparing the observed spectrum with synthesised spectra using the software  {\it Spectroscopy Made Easy, SME} \citep{sme,sme_code,sme:16}. This program  uses a grid of model atmospheres  (MARCS spherical-symmetric, LTE model-atmospheres \citep{marcs:08}) in which it interpolates for a given set of fundamental parameters of the analysed star.  The spectral lines,  which are used for the abundance analysis, are marked with masks in the pre-normalized observed spectra. SME then iteratively synthesize spectra for the searched abundances, under a scheme to minimize the $\chi^2$ when comparing with the observed spectra. In order to match our synthetic spectra with the observed ones, we also convolve the synthetic spectra with a Gaussian function of FWHM of $20\pm0.5$\,\kms . This broadening accounts for the instrumental spectral resolution and the macroturbulence  of the stellar atmosphere. 

%The best fit will provide the abundance of the element from the observed spectral lines.  In the spectral synthesis our new  accurate line list is used. 

% \textbf{LIVIA: I think that a broadening of 14 km/s is mostly instrumental. Indeed for a low mass giant macro-turbulence is normally well below 10 km/s.
% I will not quote it in Table 1 as csi-macro. I will simply mention in the text the likely instrumental broadening of 14 km/s.}

\begin{deluxetable*}{c c c c c c c c c}
\tablecaption{Stellar coordinates, position, and kinematics \label{tab:star}}
\tablewidth{0pt}
\tablehead{
\colhead{Data} & \colhead{RA} &  \colhead{dec} &  \colhead{$(l,b)$} &   \colhead{$R_c$} &  \colhead{$D_c$} & \colhead{$v^{\mathrm{helio}}_{\mathrm{rad}}$}    & \colhead{$\mu_l$}  & \colhead{$\mu_b$}  \\
   & \colhead{[h:m:s]} & \colhead{[d:m:s]} &  \colhead{[$^\circ,^\circ$]} & \colhead{[pc]} & \colhead{[pc]} & \colhead{[\kms]} & \colhead{\,mas\,yr$^{-1}$} & \colhead{\,mas\,yr$^{-1}$} 
  } 
\startdata
GC10812 & $17:45:37.229$  & $-29:00:16.62$ & ($359.9$, $-0.035$) & $1.5$ & $26$ & $-56.5$ & $2.31$ & $-3.12$  \\
Uncertainties & & & & $\pm0.1$ & $^{+54}_{-16}$ &  $\pm1.0$ & $\pm0.27$ & $\pm0.27$  \\
\enddata
\end{deluxetable*}

\begin{deluxetable*}{c c c c c c c c c c c c}
\tablecaption{Stellar photometry, parameters, and abundances \label{tab:abund}}
\tablewidth{0pt}
\tablehead{
\colhead{Data}  & \colhead{$K_S$} & \colhead{$H-K_S$} &  \colhead{$T_{\mathrm{eff}}$} & \colhead{$\log$ g} & \colhead{$\xi_{\mathrm{mic}}$} & \colhead{[Fe/H]} &   \colhead{[Mg/Fe]} & \colhead{[Si/Fe]} & \colhead{[Ca/Fe]} & \colhead{[Ti/Fe]} & \colhead{[Sc/Fe]} \\
 & & & \colhead{[K]} & \colhead{(dex)} & \colhead{[\kms]}  &   & &  & & & } 
\startdata
GC10812 & $10.25$ & $1.63$ &  $3817$ & $0.5$  & $2.0$ & $-1.05$    & $0.36$ & $0.39$ & $0.55$ & $0.53$ & $0.44$    \\
Uncertainties & $\pm0.05$ & $\pm0.07$ & $\pm150$ & $\pm0.3$ & $\pm0.5$ & $\pm0.1$  & $\pm0.1$ &  $\pm0.1$&  $\pm0.2$ &  $\pm0.3$ &  $\pm0.3$  
\enddata
\end{deluxetable*}

\begin{deluxetable*}{c c c c c c c}
\tablecaption{Uncertainties due to uncertainties in the stellar parameters\label{tab:uncert}}
\tablewidth{0pt}
\tablehead{
\colhead{parameter} &  \colhead{[Fe/H]} & \colhead{[Mg/Fe]} & \colhead{[Si/Fe]} & \colhead{[Ca/Fe]} & \colhead{[Ti/Fe]} & \colhead{[Sc/Fe]}  }
\startdata
$\Delta T_\mathrm{eff}=\pm150\,\mathrm{K}$ & $<0.02$     &$^{+0.07}_{-0.02}$ & $^{-0.04}_{+0.10}$  & $^{+0.16}_{-0.13}$  & $^{+0.28}_{-0.22}$ & $^{+0.27}_{-0.24}$     \\
$\Delta \log g =\pm0.3\,\mathrm{dex}$ & $<0.02$  & $<0.02$   & $^{+0.07}_{-0.03}$ & $^{-0.05}_{+0.17}$ & $^{-0.03}_{+0.05}$ & $^{-0.03}_{+0.05}$    \\
$\Delta \xi_\mathrm{mic} =\pm0.5$\,\kms & $^{-0.12}_{+0.10}$    & $^{+0.02}_{+0.06}$ & $^{-0.01}_{+0.05}$ & $^{-0.08}_{+0.13}$ & $^{-0.02}_{+0.06}$ & $^{-0.06}_{+0.08}$    \\
\enddata
\end{deluxetable*}

%  \begin{deluxetable*}{c c c c c c c c}
% \tablecaption{OLD Uncertainties due to uncertainties in the stellar parameters\label{tab:uncert}}
% \tablewidth{0pt}
% \tablehead{
% \colhead{parameter} &  \colhead{[Fe/H]} & \colhead{[C/Fe]} & \colhead{[Mg/Fe]} & \colhead{[Si/Fe]} & \colhead{[Ca/Fe]} & \colhead{[Ti/Fe]} & \colhead{[Sc/Fe]}  }
% \startdata
% $\Delta T_\mathrm{eff}=+100\,\mathrm{K}$ & 0  & $-0.04$ & $-0.02$ & $+0.04$ & $-0.08$ & $-0.17$ & $-0.18$    \\
% $\Delta \log g =+0.3\,\mathrm{dex}$ & 0  & $-0.12$ & $+0.06$ & $-0.12$ & $-0.01$ & $-0.04$ & $+0.03$    \\
% $\Delta \xi_\mathrm{micro} =+0.5$\,\kms & 0  & $+0.05$ & $-0.01$ & $-0.00$ & $+0.07$ & $+0.09$ & $+0.07$    \\
% \enddata
% \end{deluxetable*}
 
% \begin{figure*}[!tbp]
%  \centering
%\epsscale{1.00}
%\includegraphics[trim={0cm 1.9cm 0cm 0cm},angle=90,clip,width=0.34\hsize]{apj1.eps}\includegraphics[trim={0cm 2cm 0cm 2cm},angle=90,clip,width=0.31\hsize]{apj2.eps}
%\includegraphics[trim={0cm 2cm 0cm 2cm},angle=90,clip,width=0.31\hsize]{apj3.eps} %\includegraphics[angle=-90,width=1.2\hsize]{finding_chart4.pdf}
%\caption{Examples of spectra and uncertainties.\label{fig:feh}}
%\end{figure*}

\begin{figure}[!tbp]
  \centering
\epsscale{1.00}
\includegraphics[trim={0cm 1.9cm 0cm 0cm},angle=90,clip,width=1.05\hsize]{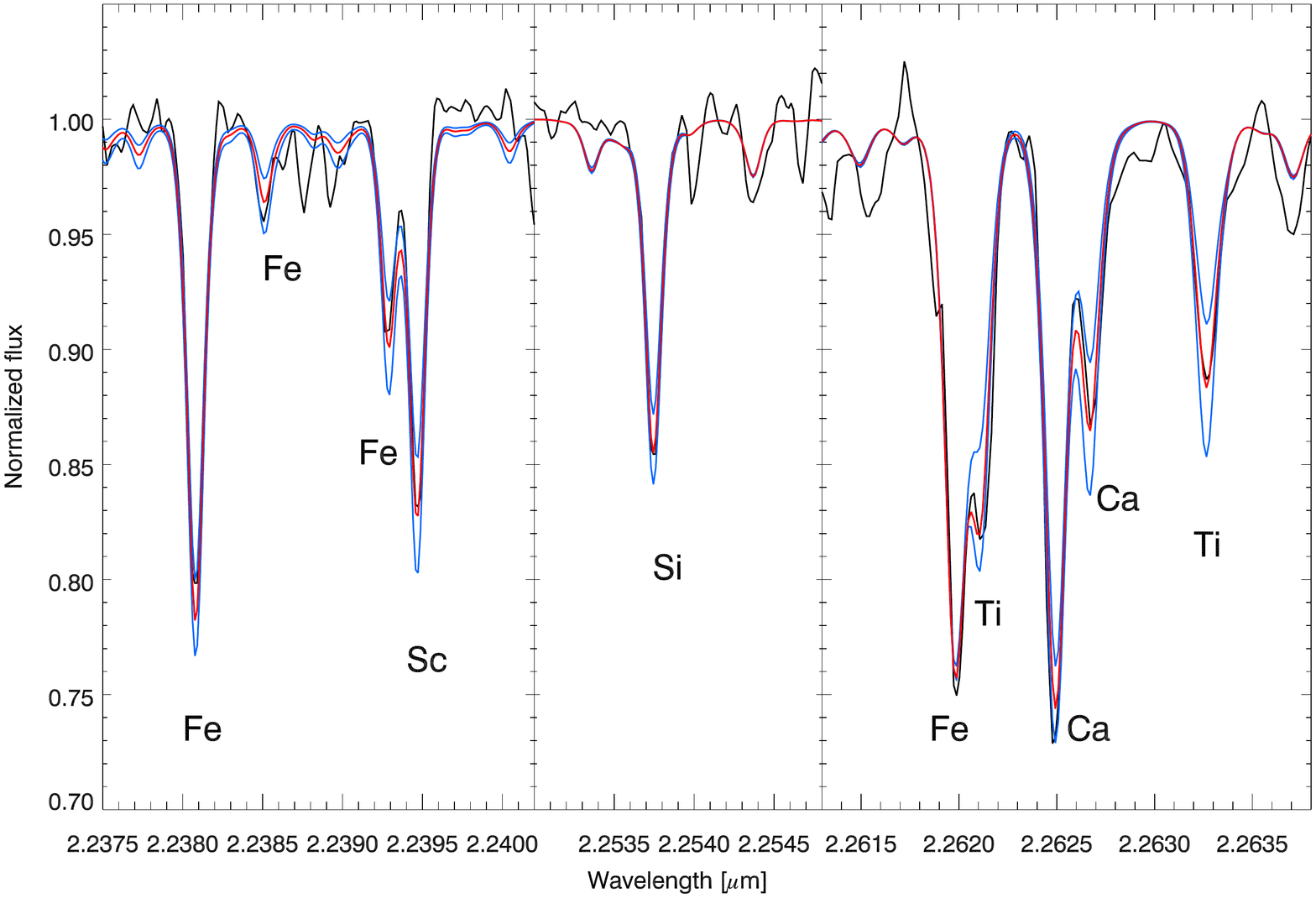} 
\caption{Examples of spectra covering a few of the lines used for the abundance determination. The black curves are the observations, the red one is the best fit model. The blue spectra correspond to $\pm0.2$ dex in the corresponding abundances, in order to show the sensitivity of these lines to the determined abundances. \label{fig:abund}}
\end{figure}

\begin{deluxetable}{c c c c}
\tablecaption{Line list\label{tab:linelist}}
\tablewidth{0pt}
\tablehead{
\colhead{Element} &  \colhead{Wavelength in air} & \colhead{Exc. Pot.} & \colhead{$\log(gf)$} \\
 & \colhead{[Å]} & \colhead{eV} & \colhead{(cgs)}}
\startdata
Mg I & $21208.106$ & $6.73$ & $-0.821$ \\
Si I & $21195.298$ & $7.29$ & $-0.425$ \\
Si I & $21779.720$ & $6.72$ & $0.418$ \\
Si I & $21819.711$ & $6.72$ & $ 0.087$ \\
Si I & $21874.199$ & $6.72$ & $-0.731$ \\
Si I & $21879.345$ & $6.72$ & $ 0.384$ \\
Si I & $22537.593$ & $6.62$ & $-0.216$ \\
Ca I & $22626.786$ & $4.68$ & $-0.281$ \\
Sc I & $21730.452$ & $1.44$ & $-1.880$ \\
Sc I & $21812.174$ & $1.43$ & $-1.490$ \\
Sc I & $21842.781$ & $1.43$ & $-1.760$ \\
Sc I & $22394.695$ & $1.43$ & $-1.180$ \\
Ti I & $22632.743$ & $1.88$ & $-2.760$ \\
Fe I & $21124.505$ & $5.33$ & $-1.647$ \\
Fe I & $21238.509$ & $4.96$ & $-1.281$ \\
Fe I & $21779.651$ & $3.64$ & $-4.298$ \\
Fe I & $21894.983$ & $6.13$ & $-0.135$ \\
Fe I & $22380.835$ & $5.03$ & $-0.409$ \\
Fe I & $22392.915$ & $5.10$ & $-1.207$ \\
Fe I & $22473.263$ & $6.12$ & $ 0.483$ \\
Fe I & $22619.873$ & $4.99$ & $-0.362$ \\
\enddata
\end{deluxetable}

%\begin{deluxetable}{c c c}
%\tablecaption{Line list\label{tab:linelist}}
%\tablewidth{0pt}
%\tablehead{
%\colhead{Element} &  \colhead{Wavelength in air} & \colhead{$\log(gf)$} \\
% & \colhead{[\AA\ ]} & \colhead{(cgs)}}
%\startdata
%Mg I & $21208.106$ & $-0.821$ \\
%Si I & $21195.298$ & $-0.425$ \\
%Si I & $21779.720$ & $0.418$ \\
%Si I & $21819.711$ & $0.087$ \\
%Si I & $21874.199$ & $-0.731$ \\
%Si I & $21879.345$ & $0.384$ \\
%Si I & $22537.593$ & $-0.216$ \\
%Ca I & $22626.786$ & $-0.281$ \\
%Sc I & $21730.452$ & $-1.880$ \\
%Sc I & $21812.174$ & $-1.490$ \\
%Sc I & $21842.781$ & $-1.760$ \\
%Sc I & $22394.695$ & $-1.180$ \\
%%Ti I & $21532.821$ & $-0.698$ \\
%%Ti I & $22443.925$ & $-2.360$ \\
%Ti I & $22632.743$ & $-2.760$ \\
%Fe I & $21124.505$ & $-1.647$ \\
%Fe I & $21238.509$ & $-1.281$ \\
%Fe I & $21779.651$ & $-4.298$ \\
%Fe I & $21894.983$ & $-0.135$ \\
%Fe I & $22380.835$ & $-0.409$ \\
%Fe I & $22392.915$ & $-1.207$ \\
%Fe I & $22473.263$ & $0.483$ \\
%Fe I & $22619.873$ & $-0.362$ \\
%\enddata
%\end{deluxetable}
%

The abundances of the elements Fe, Mg, Si, Ca, Ti, and Sc are determined from carefully chosen lines. We have restricted our analysis to lines that are on the weak part of the curve of growth, i.e. with equivalent widths of $W<250$\,\AA\ or $\log W/\lambda < -4.9$. This means that at our resolution of $R=24,000$, there is an upper limit to the line depth of approximately 0.75 of the continuum. Lines deeper than this will certainly be saturated and will not be as sensitive to the abundance but at the same time more sensitive to the uncertain microturbulence parameters, $\xi_{\mathrm{mic}}$. Our restriction of the analysis to weak lines will ensure a good measurement of the abundances. We also require that the spectral recording around the lines and the form of the continuum is of such high quality that the continuum is traceable. 

The final line list is given in Table \ref{tab:linelist}. In the Table the wavelengths, excitation potential, and the line strengths of the lines used for the abundance determination are given. The entire line list including all lines will be published elsewhere (Thorsbro et al. 2017, in prep.).

 Examples of synthetic spectra are shown in Figure \ref{fig:abund} and the derived abundances are given in Table \ref{tab:abund}. 
The Fe, Si, and Sc abundances are determined from 8, 6, and 4 lines, respectively, whereas the Mg, Ca, and Ti abundances are determined from only one carefully chosen line each. 
This means that the former abundances are observationally better determined.
%carbon abundance is determined from 5 weak CO lines and the 
%Fe, Mg, Si, Ca, Ti, and Sc abundances are determined from a 2-6 lines for every element. 

The uncertainties in the determination of the abundance ratios, for typical uncertainties in the stellar parameters, are $\pm0.1$ for [Fe/H], [Mg/Fe], and [Si/Fe], whereas it is $\pm0.2$\,dex for [Ca/Fe] and $\pm0.3$\,dex for [Ti/Fe] and [Sc/Fe], see Table \ref{tab:uncert}. In addition, we estimate random uncertainties of less than 0.05 dex due to the continuum placement.
%The latter we estimate to of the order of 0.10 dex. Adding the uncertainties in quadrature will overestimate the total uncertainties. We therefore estimate the total uncertainties in the derived abundances as being of the order of $0.1-0.3$ dex, and these are given for each element in the second row in Table \ref{tab:abund}. 
% {\color{blue}Tobias: I am guessing that the main reason for the error are the $T_{eff}$ and so uncertainties and the statistical uncertainties are irrelevant. Also I assume that one has not to the errors from Table 3 to the error of Table 2. However, that is maybe not obvious for everyone, maybe write that explicit.}
In Figure \ref{fig:abund} we  show the observed and our best fit  spectra, and also synthetic spectra with $\pm0.2$\,dex abundance variations with respect to the best fit solution. The Figure clearly shows that we can derive abundances with overall uncertainties smaller than $0.2$\,dex.

% {\color{green} Nils: 
% Vi har fått logg med photometry.
% Kan vi få ut logg från våra spektra? plotta en model med logg=3.9?   Use tuans teff.
% Kan ej vara dvärg. Visa. Kan Mathias få ut logg från NaI? }

\begin{figure*}[!tbp]
  \centering
\epsscale{1.00}
\includegraphics[trim={0.00cm 0.0cm 0.0cm 0.0cm},angle=90,clip,width=1\hsize]{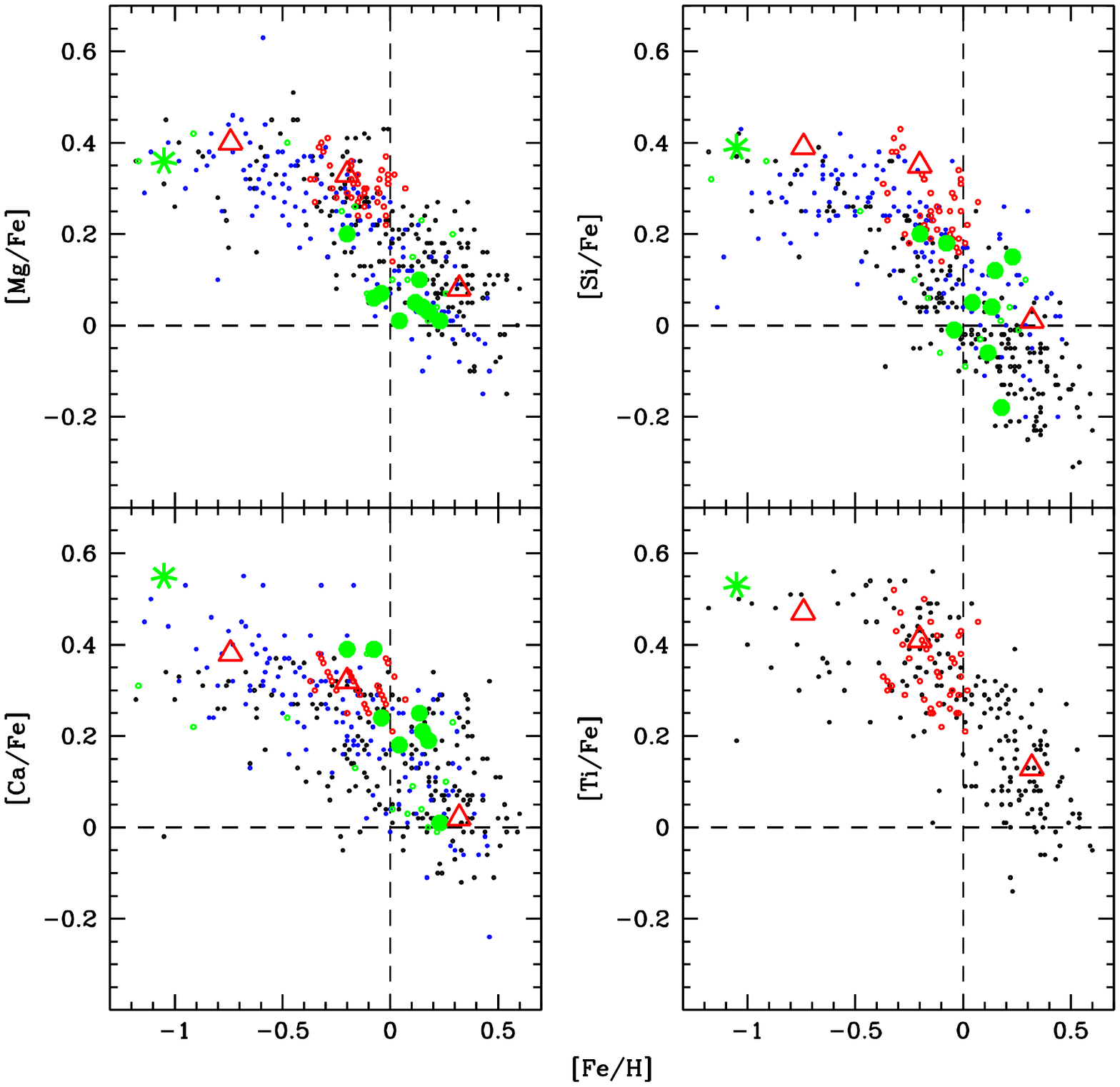} 
\caption{Abundances ratios of [Mg/Fe], [Si/Fe], [Ca/Fe], and [Ti/Fe] versus [Fe/H] for  GC10812 (big green star) 
and different samples of bulge and Galactic Center giants: \citet{gonzalez:11} (black dots), \citet{johnson:2014} 
(blue dots) in the outer bulge from optical spectroscopy; \citet{rich:07,rich:12} (red circles) and \citet{ryde:16} 
(green circles) in low latitude bulge fields and \citet{ryde_schultheis:15}  (big green dots) in the Galactic Center region from IR spectroscopy. The average abundance ratios of the three stellar populations of Terzan~5 from \citet{origlia:11,origlia:13}
(big red triangles) are also plotted for comparison.
 We have re-scaled all the abundances to the same solar reference of \citet{solar:sme}. \label{fig:afeh}}
\end{figure*}

\section{Results}
\label{sect:results}

\begin{figure}[!tbp]
  \centering
\epsscale{1.00}
\includegraphics[trim={0cm 0cm 0cm 0cm},angle=0,clip,width=1\hsize]{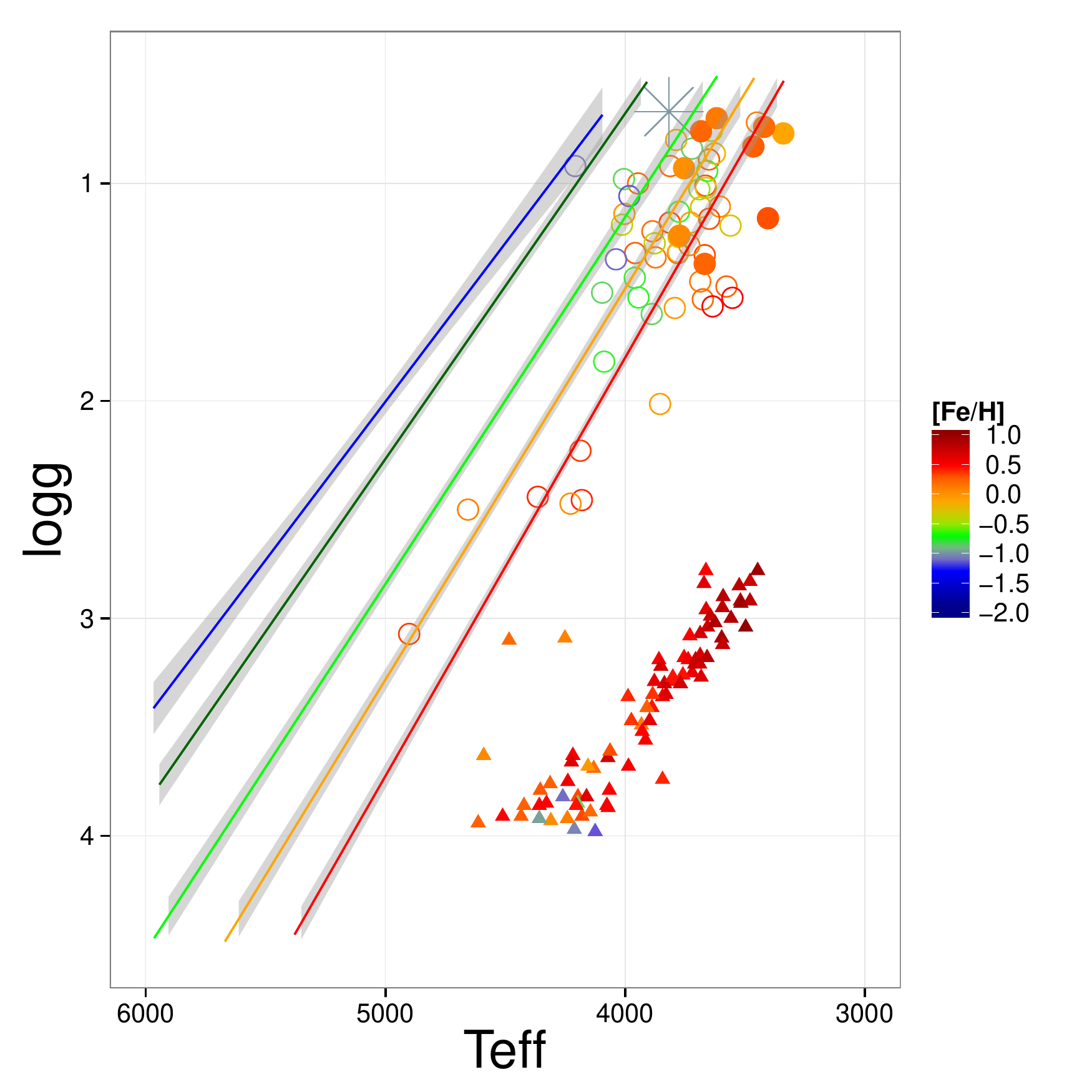} 
\caption{Logarithm of the surface gravities of inner bulge stars plotted versus their effective temperatures and color-coded by their metallicities. The big star is the metal-poor star GC10812 discussed in this paper.  The selection of stars in the inner $3^\circ$ of the bulge from the literature are marked by open circles for the stars from \citet{ryde:16,schultheis:15}, by filled circles for the stars within 10 pc projected galactocentric distance from \citet{ryde_schultheis:15}, and by triangles for the stars discussed in \citet{do:15}.  Superimposed are also the 10 Gyr  PARSEC isochrones \citep{bressan:12} with the corresponding color of the metallicity. 
\label{fig:cmd}}
\end{figure}

%Livia IS updating the section with the new Figure 3
Our derived stellar parameters and their uncertainties for GC10812 are given in Table\,\ref{tab:abund}. We find that these parameters place the star in the appropriate location on the Hertzsprung-Russel diagram, as seen in Figure \ref{fig:cmd}. It lies on the red giant branch appropriate for its metallicity, as demonstrated by plotting an old population with the 10\,Gyr isochrones of \citet{bressan:12}.

Our derived abundances of Fe, Mg, Si, Ca, Ti and Sc are given in Table \ref{tab:abund}. We have normalized our derived abundances to the solar abundances of \citet{solar:sme}: 
$\mathrm { \log\varepsilon(Mg)=7.53}$, $\mathrm { \log\varepsilon(Si)=7.51}$, $\mathrm { \log\varepsilon(Ca)=6.31}$,
 $\mathrm { \log\varepsilon(Sc)=3.17}$, $\mathrm { \log\varepsilon(Ti)=5.02}$,  and $\mathrm { \log\varepsilon(Fe)=7.45}$.

Typical internal errors in the derived stellar abundances are a few hundredths dex, while the systematic uncertainties due to different assumptions for the stellar parameters are detailed in Table \ref{tab:uncert} and, on average, amount to 0.1-0.2 dex, often dominated by one of the uncertainties in the stellar parameters.

The derived [$\alpha$/Fe] abundance ratios are plotted in Figure \ref{fig:afeh}, together with the corresponding measurements of different samples of giants in the bulge from \citet{gonzalez:11} and \citet{johnson:2014} by means of optical spectroscopy, as well as measurements in low latitude (innermost 2 degrees) fields from \citet{rich:07,rich:12} and \citet{ryde:16} by using H and K band IR spectroscopy.
In Figure \ref{fig:afeh} we also reported the measurements of some low mass giants in the Galactic Center region from \citet{ryde_schultheis:15} and the average abundance ratios of the three stellar populations of Terzan~5 at $(l,b)=(3.8^\circ$,$+1.7^\circ)$ from \citet{origlia:11,origlia:13}.

We find that the [$\alpha$/Fe] abundance ratios of GC10812 are consistent with an enhancement between a factor of two and three with respect to the solar values and fully consistent the values measured in bulge and Galactic Center giants with sub-solar metallicities.

\section{Discussion}

The best constraint of the 3D location of GC10812 is at $1.5$\,pc distance from the SgrA* in projected distance on the sky and $26^{+54}_{-16}$\,pc distance in the line-of-sight in front of the Galactic Center.  The nuclear cluster has a half-light radius of $178\pm 51" \sim 7\pm 2$\,pc \citep{fritz16} but the nuclear disk extends to about 230 pc \citep{launhardt02}. The dynamics of the star is typical for a Galactic Center star  and our orbit calculations show that the star's orbit is constrained within the nuclear disk. 

 The kinematics strongly  favor a galactic center origin
and it is thus rather certain that the star belongs to a nuclear component. The probability of the star being a stray bulge giant is low since most bulge giants are more metal-rich and [Fe/H]$\sim-1$ stars are rare in the bulge \citep{ness:16}. Based on the kinematics and metallicity,
the probability of the star being a halo giant is also low
since at least in the solar vicinity the halo metallicity in
the mean is typically [Fe/H]$\sim-1.6$ not $-1$  \citep{ryan:91}. The relative number density of halo stars in
the Galactic Center is difficult to estimate, since there are
no measurements of the halo quantitatively even within 8
kpc. Extrapolation can obtain a relevant mass in the center when a broken power-law like, for example, that in \citet{bland16} is used. However, when other parametrizations are used, like an Einasto profile \citep{sesar:13,xue:15},  the obtained mass is much smaller and irrelevant. Also from a theoretical standpoint a flatter profile is expected, since it is difficult for dwarf galaxies, which probably form the main contributor to the halo, to reach the center of the Milky Way. For example \citet{bullock:05} measure, in the their model, a slope of $-1$ between $3$ and $10$~kpc, flatter than further out. 
In summary,
we conclude that the probability of GC10812 being a halo
star is low.

The metallicity of  GC10812 is lower than of most stars in inner bulge. \citet{rich:07,rich:12} find, for example,  a dispersion of approximately 0.1 dex around [Fe/H]=$-0.05$ to $-0.15$. The more recent works of \citet{schultheis:15,ryde:16} found in addition also some metal poor stars with [Fe/H]$\approx-1$. They are with $|b|>0.4^{\circ}$ all located outside the nuclear disk. Within the nuclear components (disk and cluster)  \citet{cunha2007} find a total spread of $0.16$ dex around [Fe/H]=$+0.14$ for the luminous giants and supergiants located within $2.2$\,pc of the Galactic Center. Further, 
\citealt{carr2000}, \citealt{ramirez:00}, and \citealt{davies2009} analyzed high-resolution spectra  of supergiant stars in the Galactic Center finding near-solar metallicity. Similarity \citet{ryde_schultheis:15} at about 4 arcmin distance from Sgr~A* also only detect a metal-rich component with a total spread of $0.15$ dex around [Fe/H]=$+0.11$.

The location of the  [$\alpha$/Fe] abundance ratios versus metallicity for the giant GC10812 is that expected for the metal-poor population in the outer bulge.  The [Mg/Fe] and [Si/Fe] trends of the inner bulge are tight and are indistinguishable from the outer bulge trend, within uncertainties. Although with a larger scatter, this is also true for the [Ca/Fe] abundances for giants in the central regions \citep{cunha2007,origlia:11,johnson:2014,ryde:16}. It can, however, be noted that they all follow a higher trend than that determined by \citet{bensby:13}. This might (but not necessarily) arise from greater errors ultimately attributable to the higher uncertainties based on the Ca determination based on the giant star spectra as compared with those based on dwarf spectra \citep[see also the discussion in][]{gonzalez:11}. Scrutinizing the Ca line used, using more Ca lines, comparing with detailed galactic chemical evolution models, and observing more stars will be needed to investigate and understand the true nature of the [Ca/Fe] trend. Our [Ti/Fe] determination also has a high uncertainty, mainly arising from the uncertainty in the effective temperature. The value is, however, within errors consistent with the other bulge stars too. 
Thus, within the uncertainties, the [$\alpha$/Fe] we measure for GC10812, cannot be claimed to be different to the rest of the bulge.
% Only a decrease in temperature within its uncertainties would decrease the Ti abundance to 0.31, [Ca/Fe]=0.42, [Si/Fe]=0.49, at the same time as keeping the [Mg/Fe] and the metallicity unchanged. The uncertainties are certainly correlated, but this shows that, within the uncertainties the [$\alpha$/Fe] we measure for GC10812, cannot be claimed to be different from the rest of the bulge.

The $\log g$ versus $T_\mathrm{eff}$ location of GC10812  is indicated in Figure \ref{fig:cmd} with a large asterisk. Superimposed are isochrones color-coded for different metallicities \citep{bressan:12}. We assume here an age of 10\,Gyr. The red line shows the most metal-rich isochrone ($+0.7$\,dex). Based on its apparent luminosity, its kinematic membership to one of the nulcear components, its low metallicity, our  independent determination of its effective temperature and surface gravity, and its high $\alpha$-abundance, we are confident that GC10812 is consistent with being a low-mass, old red-giant star in the vicinity of the nuclear cluster. 
% {\color{blue}Tobias: can we change the age to 12 Gyrs? I think the main assumption now for the majority of the blue and would decrease the disagreement for very metall rich stars.} 

In the Figure we also plot the locations for a sample of stars in the inner $3^\circ$ from the Galactic Center from \citet{ryde_schultheis:15}, \citet{ryde:16}, \citet{schultheis:15}, and \citet{do:15}.
The typical uncertainties of temperatures are about $\pm 150$\,K while the errors in
$\log g$ can be  in the order of 0.3-0.5\,dex as those were determined photometrically. We note that the isochrones  predict too high temperatures, for a given surface gravity, for the most metal-rich stars ($\rm > 0.5\,\mathrm{dex}$).  One should, however, be aware of the fact that everything at metallicity $>+0.5$\,dex needs still to be understood: the metallicities themselves, model atmospheres, and isochrones are all very uncertain and mostly calibrated by extrapolation.
Thus, it could be possible that the $\log g$ determination of the most metal-rich stars are $\sim 0.3-0.5\,$dex too high.
% what we know at about solar. I will not bet on any quantity and relative uncertainty in this regime.}

Whereas the M giants from \citet{ryde_schultheis:15}, \citet{ryde:16} and \citet{schultheis:15} are situated along the isochrone sequence in Figure \ref{fig:cmd}, the location  of the \citet{do:15} stars is not compatible with the indicated location of the RGB branch from the PARSEC isochrones. These stars are plotted as triangles. As their extinction is typically that of stars in the Galactic Center, we believe that their surface gravities are about two to three order of magnitudes too high.  
% The surface gravities are more consistent with dwarf stars, which cannot be correct if the stars are located in the nuclear cluster. Nor can all these stars be foreground dwarfs in such a small cone towards the galactic center. 
Also, the metal-poor stars discussed in \citet{do:15}, are orders of magnitude away from the expected isochrones. 
%Obviously, these stars as dwarfs and cannot be lying in the Galactic center. {\color{blue}Tobias: it is not possible that they are dwarfs, because the extinction matches at least roughly the GC extinction, thus there are giants. Also the shape in the diagram is like a giant branch with a wrong calibration. For me the offset is usually about 2 orders of magnitude not 3.} If on the other hand these dwarfs are actually giants in the Galactic Center, the gravities might
%be off by as much as three orders of magnitude. 

The work of \citet{do:15} obtained integral-field, moderate-resolution spectroscopy for scores of stars in the central cluster behind AO.   They did not claim to undertake a full high-resolution abundance analysis and were aware of potentially significant uncertainties in their methods.    Their stated uncertainty in $\log g$ is $0.91$\,dex.  The main effect of decreasing the surface gravity in a synthetic spectrum calculation is the decreased continuous opacity, which generally increases line strengths. In the simultaneous fit of the stellar parameters by \citet{do:15}, the temperature  and metallicity determinations might  therefore also be affected by this large uncertainty. 
%also be off by a large amount. Since the metallicity is a function of $\log g$ and $T_\mathrm{eff}$, their metallicities are affected by an unknown amount. 
 We agree with the assessment of  \citet{do:15} that additional observations at high spectral resolution would be required to confirm the low metallicities ([Fe/H]$\sim -1$) claimed for the five stars in the nuclear cluster.
%A reanalysis, with higher resolution spectra, of the five stars in the nuclear cluster claimed to be of low metallicities ([Fe/H]$\sim -1$) would be necessary to confirm the existence of metal-poor stars there. 
Likewise, the high-metallicity stars found by  \citet{do:15}, which have nominal metallicities up to [Fe/H]$=+1.0$\,dex, need to be confirmed.
%, by redetermining the stellar parameters.  

% {\color{blue}Tobias: What causes the $T_{eff}$ error? Systematics or statistics? Can it maybe improved with some changes because Ti looks interesting, but is currently limited by it.}

%... not globular clusters that have evaporated to form the cluster.

%globular clusters that have evaporated to form the cluster.
%More on different formation scenarios of the nuclear clus-
%ter??: infall of GC or insitu SF from cosmological gas in-
%flow.... Most stars are of solar metallicity, indicative of %in an
%in situ origin.

%%The low metallicity of GC 10812 could indicate an origin in globular clusters... of Bulge globular clusters [Fe/H]$\sim 0.75$...}

The similarity between GC10812 and the rest of the inner bulge,  would suggest a homogeneous star-formation history in the entire bulge. There is a clear connection with the bulge and the Center.
Thus, our results argues for the Center being in the context of the bulge over most of its history rather than very distinct.
%Tobias: new, but I'm not sure whether that is really helpful

% {\color{red} Still, the the present epoch existence of gas, star formation, and young massive stars in the center attests to a clear difference. Possibly, that difference was only introduced recently, perhaps by the bar. That would argue that the bar is especially strong today. Another possibility might be that such star-formation episodes happen sometimes for a short period followed by a larger quiescent period and that, just by incident, we live in a period with star formation in the center. Such a cycle could also be caused by the bar. 
% In any case, it seems that this recent star formation is small compared to the star formation 10 Gyrs ago \citep{pfuhl:11}  and that 
% there may have been little difference between the very center and the bulge then. }

% \begin{figure}[!tbp]
%   \centering
% \epsscale{1.00}
% \includegraphics[trim={0cm 0cm 0cm 0cm},angle=0,clip,width=1\hsize]{Kvslogg.pdf} 
% \caption{K vs logg \label{fig:cmdK}}
% \end{figure}

% {\color{blue}Tobias: do we know the age of the stars? it should be possible I assume since we know the distance, still there should be an age error error. How large is the error? The age could be important for origin interpretation.}
%Or can be use as APOGEE CN for ages?}

\section{Conclusions}

In targeting the Milky Way nuclear star cluster we have observed the most metal-poor giant (GC10812) yet in the vicinity of the Galactic Center. 
A careful analysis of its 3-dimensional location, locates it at $26^{+54}_{-16}$\,pc in front of the Galactic Center and at a projected distance of 1.5 pc to the North-West. This line of sight position and orbit integration makes it unlikely that the star belongs the most central component, the nuclear cluster. However, the orbit integration also shows that the star very likely does not leave the nuclear disk. Thus, the star very likely belongs to a nuclear component.

The metallicity and abundances are determined from a detailed abundance analysis based on $R=24000$ Keck/NIRSPEC spectra. The [Fe/H]$=-1.05\pm0.10$ is the lowest measured and confirmed metallcitiy of a star from the nuclear components. It is unexpected and differs from earlier measurements. We can, however, still conclude that there is no evidence hitherto that there
are metal-poor stars (e.g. originating in globular cluster in-spiraling to the Galactic Center \citep{tremaine75,capuzzo08}) % not much now, because star is probably not a member of the nuclear cluster
in the nuclear star cluster. %We have discovered a metal-poor field star in the vicinity of it 
The [$\alpha$/Fe]-element enhancement of $\sim+0.4$ follows the trend of the outer bulge. 

GC10812 is by virtue of its 3D kinematics a likely member of the central disk/cluster system.  It also exhibits the metal poor, alpha enhanced hallmarks of an old, metal poor giant.  The existence of an old population in the Galactic Center has been well established from the robust presence of a red clump population \citep{figer}  as well as the analysis of star formation history by  \citet{pfuhl:11}.
It will be important going forward to explore the full abundance distribution of this old population, as well as that of the $\sim 10^8$\,yr population responsible for the supergiants.   Such studies will lay the foundation for applying models of chemical evolution to this very interesting region of the Milky Way.

%Figer, D. F., Rich, R. M., Kim, S. S., Morris, M., &Serabyn, E. 2004, ApJ, 601, 319

% We have demonstrated that the chemical description of the nuclear cluster will be possible to retrieve based on K-band spectroscopy at high spectral resolution. This will infer the system's stars-formation and enrichment history as well as its relationship with the central 100 pc of the bulge/bar system. Spectra of more stars and a more quantitative analysis of metallicity distributions are necessary to constrain these further.

We have demonstrated that K band spectroscopy of individual giants at high spectral resolution offers a path forward enabling exploration of the chemistry of the central cluster.  This has the potential to elucidate the system's star formation and enrichment history as well as its relationship with the central 100 pc of the bulge/bar system.

\acknowledgments
We would like to thank the referee for an insightful and careful report which improved the paper. Nikolai Piskunov is thanked for developing the spectral synthesis code, SME, to handle fully spherical-symmetric problems. Asli Pehlivan is thanked for providing atomic data on Mg lines prior to its publication. N.R. acknowledges support from the Swedish Research
Council, VR (project number 621-2014-5640), and Funds from Kungl. Fysiografiska
Sällskapet i Lund (Stiftelsen Walter Gyllenbergs fond and Märta och
Erik Holmbergs donation). R.M.R. acknowledges support from  grant AST-1413755 from the US National Science Foundation. L.O. acknowledges PRIN INAF 2014 - CRA 1.05.01.94.11:  "Probing the internal dynamics of globular clusters. The first, comprehensive radial mapping of individual star kinematics with
the new generation of multi-object spectrographs” (PI: L. Origlia).  S.C. acknowledges support from the Research Center for Astronomy, Academy of Athens. The authors wish to
recognize and acknowledge the very significant cultural role and 
reverence that the summit of Mauna Kea has always had within the
indigenous Hawaiian community.  We are most fortunate to have the
opportunity to  conduct observations from this mountain.

%\appendix
%\section{appendix section}

%\bibliographystyle{yahapj}
\bibliographystyle{apj}
%\bibliography{references}

\end{document}